# PRIVACYCUBE: DATA PHYSICALIZATION FOR ENHANCING PRIVACY AWARENESS IN IOT

**Bayan Al Muhander**
School of Computer Science and Informatics
Cardiff University, UK
almuhanderb@cardiff.ac.uk

**Nalin Arachchilage**
School of Computer Science and Informatics
University of Auckland, New Zealand
nalin.arachchilage@auckland.ac.nz

**Yasar Majib**
School of Computer Science and Informatics
Cardiff University, UK
majiby@cardiff.ac.uk

**Mohammed Alosaimi**
School of Computer Science and Informatics
Cardiff University, UK
alosaimimm1@cardiff.ac.uk

**Omer Rana**
School of Computer Science and Informatics
Cardiff University, UK
RanaOF@cardiff.ac.uk

**Charith Perera**
School of Computer Science and Informatics
Cardiff University, UK
pererac@cardiff.ac.uk

## ABSTRACT

People are increasingly bringing Internet of Things (IoT) devices into their homes without understanding how their data is gathered, processed, and used. We describe PrivacyCube, a novel data physicalization designed to increase privacy awareness within smart home environments. PrivacyCube visualizes IoT data consumption by displaying privacy-related notices. PrivacyCube aims to assist smart home occupants to (i) understand their data privacy better and (ii) have conversations around data management practices of IoT devices used within their homes. Using PrivacyCube, households can learn and make informed privacy decisions collectively. To evaluate PrivacyCube, we used multiple research methods throughout the different stages of design. We first conducted a focus group study in two stages with six participants to compare PrivacyCube to text and state-of-the-art privacy policies. We then deployed PrivacyCube in a 14-day-long field study with eight households. Our results show that PrivacyCube helps home occupants comprehend IoT privacy better with significantly increased privacy awareness at p < .05 (p=0.00041, t= -5.57). Participants preferred PrivacyCube over text privacy policies because it was comprehensive and easier to use. PrivacyCube and Privacy Label, a state-of-the-art approach, both received positive reviews from participants, with PrivacyCube being preferred for its interactivity and ability to encourage conversations. PrivacyCube was also considered by home occupants as a *piece of home furniture*, encouraging them to socialize and discuss IoT privacy implications using this device. Watch the demo (Demo Video) (Source Code).

***Keywords*** Privacy Awareness; Physical Visualization; Usable Privacy; Design Space

## 1 Introduction

Numerous smart home IoT devices can collect massive amounts of data and pose privacy risks [1]. Yet, users have minimal ways to learn about data practices adopted by these devices [1, 2, 3, 4, 5]. Personal data protection regulations, such as the General Data Protection Regulation (GDPR) [6] and the California Consumer Privacy Act (CCPA) [7], enforce transparency while collecting and processing personal data. In GDPR, data controllers (i.e., entities collecting users' data) must provide users with privacy notices informing them about current or potential data processing practices.

However, available notices have been largely ignored, abandoned, and often forgotten over time [8]. Furthermore, Existing notices' designs often convey little information about the system's data practices or are website-specific and not IoT-focused [9]. Many available notices also only display one type of notice with no variation based on device/data type and use.

Despite the increasing emphasis on designing usable privacy interfaces, current designs often fail to provide users with comprehensive and contextually relevant information about data practices [9]. Privacy studies have found that privacy notices often fail due to their presentation, as most of them are long and difficult to read [10]. Furthermore, the limited ability of these interfaces to engage users exacerbates the problem of privacy management on IoT devices.

Several researchers explained that people could understand privacy implications, engage in socially meaningful actions, and make informed privacy decisions through better privacy notice presentations [11, 2]. Privacy notices and their effectiveness also depend on the physical display on which they are presented to a user. Unlike mobile apps, physical displays have long been used for learning and exploration as they are a more visually appealing embodiment of complicated concepts [12]. In particular, physical *cubes* have visualization properties that demonstrate their effectiveness in attracting user attention in a home environment [13]. Creating contextualized, efficient, and simple privacy user interfaces is still beyond the capabilities of commercial IoT products. Hence, this paper is **motivated** by the need for creativity and development to create effective privacy interfaces tailored to the IoT environment's unique demands.

We present PrivacyCube (Figure 2), a novel interactive data physicalization designed to raise privacy awareness in smart homes. PrivacyCube displays privacy notices with more meaningful information for the user. PrivacyCube can display variable privacy-related notices based on IoT devices within a home environment and the data such devices collect. PrivacyCube engages users through physical interactions, such as rotation, touch, placement in space, and interaction with embedded lights and icons.The privacy notices displayed on PrivacyCube are mostly based on a privacy infrastructure created by Das et al. [14]. The goal is to provide households with an easy way to view and understand the privacy aspects of multiple IoT devices simultaneously, to prompt them to evaluate and review data management practices. Consequently, their ability to make informed privacy decisions will improve.

We monitored the devices' traffic in order to trigger PrivacyCube's notifications. This traffic was utilized for the sole purpose of generating the notifications. Once we discover that the device is active, the code will map the device to its pre-classified characteristics, which we retrieved from the device's privacy policies. The cube will then turn on the associated LEDs. We want to emphasize that our research was not centered on deducing the activities of encrypted devices from traffic data, as this is out of this research scope. Instead, we sought to visualize privacy-related information contained in IoT devices' privacy policies and provide a more usable interface for users. PrivacyCube consists of several modules detailed in Appendix A.

PrivacyCube has five faces, with the sixth serving as the base. The cube's top face has a touch screen that displays the IoT resources. On the other faces, there are multiple LEDs showing five privacy factors based on [15]: type, storage location and retention, usage, and access. Figure 2 depicts the design of PrivacyCube, motivated by the need to provide users with a physical privacy notice. We suggest this will improve user awareness of the processes performed on their data and enable them to make appropriate privacy decisions. Deploying PrivacyCube will also help users understand the capabilities of IoT devices. The results of both focus groups and the field study suggested that PrivacyCube provides a more intuitive and user-friendly way that effectively transmits privacy information than non-physical interfaces. This work provides the following research contributions:

- Privacy policies and notices include critical information concerning collecting and processing consumer data; unfortunately, the vast majority of privacy policies and notices are largely ignored. To date, most privacy policy usability research has been conducted or requires the use of Web or mobile applications. **To the best of our knowledge, this is the first study to establish a link between IoT privacy policies and physical objects at the scale of a smart home environment.**

- **Implementation of PrivacyCube:** We develop a product-like and aesthetically neutral design that can be used in real-world settings without affecting everyday behaviour. We used a physical cube to visualize smart home IoT data consumption using embedded LEDs and an interactive screen. Through various colored physical notifications, users can explore, learn and engage with privacy policy data.

- **Evaluation of PrivacyCube:** We conducted two evaluations to determine the usability of our design. A two-phase focus group with six participants to discuss design usability and a two-week field study with eight smart home households at various home locations to demonstrate that PrivacyCube can be used in practical application scenarios. The study resulted in insightful and improved privacy awareness, and participants rated PrivacyCube as an effective tool for learning about IoT data privacy.

- **Exploration of interesting usage contexts:** PrivacyCube opens new possibilities for users to interact with their data and engage in privacy-related conversations. We show how PrivacyCube has the potential to improve privacy awareness and encourage individuals to make informed privacy decisions.

This paper is structured as follows: Section 2 describes related work. Section 3 describes the design and implementation of PrivacyCube. PrivacyCube's evaluation using a two-phase focus group study and a two-week field study is described in Section 4. Section 5 discusses PrivacyCube's applicability, limitations, and future research opportunities. followed by the conclusion in Section 6.

## 2 Related Work and Motivation

### 2.1 Privacy Policy Visualization

Current efforts in usable privacy notices examine how to effectively communicate privacy information to users through different design elements [15]. Shortening privacy policies and using simpler and neutral-colored icons and text has been proposed in [3, 16, 17]. Some studies have added interactive elements to their visualization, such as notification bubbles [18], pop-up [19] and hovering elements [20], where the user can enquire for more information. Other studies explored visualizing policies as labels presenting the types and quantities of collected data, as well as the retention and disclosure principle [21, 22]. Grid views [23, 24], and circular or wheel view [20, 25, 26] were further proposed to visually organize privacy policy and to encourage users to read them.

Most of the above initiatives have mainly focused on Web and mobile privacy, with limited focus on IoT [15]. The nature of the IoT often exacerbates many challenges associated with ensuring data subjects understand their data, such as IoT heterogeneity and variable collected data types [27]. We extend existing research by developing a data physicalization to display privacy notices. As reported in the literature, physical representations convey meaning and emotion in data [28], which can help people think, understand and engage with their data [29]. PrivacyCube's physical properties can encourage learning and exploration about IoT privacy.

### 2.2 Physical Interface

Recent research identified the key components of effective privacy notices, which can inform users about the processes carried out on their data [21, 30, 8, 31]. Promising attention is given to physical formats as they increase interactivity and engagement [32, 13, 29]. Physical devices have been used as a privacy notice delivery method, where a physical device, such as wearable or smart appliance, communicates a warning message about possible data usage [33, 29]. Several studies, including [34, 35, 36], specified that many individuals do not consider the implication of IoT devices on their lives since these implications are usually invisible, i.e., in a nonphysical format. The individuals surveyed in the above studies show more respect for physical contact with objects within their sight and preferred objects they can touch and see.

We consequently created PrivacyCube as a data physicalization to display privacy notices. We chose the cube shape because it has more than one face. Face diversity provides us more space to include the crucial aspects of a privacy notice without overwhelming the user with lengthy details [37]. The cube shape was previously used as an interface to deliver information to the user in [13, 38]. According to research, cube formats and physical interfaces have unique properties, like natural affordance [37], that support increasing individual awareness [39, 12]. Another significant advantage of data physicalization is that it is direct, intuitive, and unobtrusive information delivery technique. Literature has also emphasized the benefits of an individual's interactions with physical objects [40, 41, 39]. Data physicalization has distinct properties, such as being close to individuals' attention and allowing them to use their natural interaction skills with physical objects [42]. The previous properties, as cited in [40, 43, 44], can help users understand and improve their trust in IoT.

### 2.3 IoT in Smart Homes

In a smart home setting, each smart home sensor has its own privacy policy, i.e., a person would have to read multiple privacy policies to comprehend the processes performed on their data, resulting in a "problem of scale" [45]. For instance, the average internet user spends around 244 hours reading privacy policies per year [46], which, if applied to IoT devices, can exceed the amount of time spent using the device, hence appropriate privacy notification mechanisms are required. The notification methods must be designed to be comfortable for individuals to incorporate into their homes.

Surveys reveal that individuals consider privacy and the difficulty of privacy-related decision-making as top concerns when using IoT [47, 48, 49, 50, 51]. Users are usually unaware of the collection and processing of their data in their homes because data collection is often non-transparent and not visible to them. Prior studies addressing this issue discussed adding physical locators to IoT devices to help users become more aware of nearby devices [52, 53]. However, this solution only informs users about the existence of an IoT device without providing information about data usage, access, and retention. Our proposed PrivacyCube uses data physicalization to alert users to the presence of an IoT device. PrivacyCube's neutral and aesthetic design allows it to blend in as a piece of furniture in a home environment. Furthermore, PrivacyCube displays privacy-related information for each IoT device, allowing users to be more aware of their data.

## 2.4 Motivation

The decision to employ data physicalization, exemplified by PrivacyCube, is grounded in their inherent benefits to user comprehension and engagement. They serve as a bridge between digital and physical realms, which can enhance understanding by enabling users to interact physically with abstract privacy concepts. Emphasizing their capacity suggests that they can be tailored to household routines through fun ways to discover, learn, and share complex concepts [54]. Unlike non-physical interfaces, PrivacyCube, as evidenced by the results from both focus groups and the field study, offers a more intuitive and user-friendly interface for communicating privacy information. We envision that PrivacyCube can provide a valuable solution to the complexities of conveying privacy in the IoT context.

## 3 PrivacyCube: Design Principles

The design of our privacy policy visualization tool is informed through existing literature [17, 20], privacy professionals, and product designers. After careful consideration of the design space [55, 17], we used the following design principles to inform the design of PrivacyCube.

- **Comprehensive:** Ideally, the tool should display the complete data practices as described in current *traditional* text-based privacy policies.
- **Usable:** The tool should display meaningful privacy information and enable users to interpret information quickly, accurately, and confidently, even if they have never used it before. It should require minimal effort.
- **User Awareness:** The tool should simplify learning about active IoT devices and their data usage practices. Users should have a positive experience using the tool to enhance their privacy awareness.
- **Neutrality:** The tool should provide users with a neutral ground to interact with their privacy notifications.

## 3.1 Comprehensive

To decide on the content of a comprehensive privacy policy, we primarily referred to the P3P protocol [56, 57] and the privacy infrastructure by Das et al. [14], from which we concluded the main factors required in an effective privacy notice. These factors have also been identified in several studies as the primary causes of privacy concerns raised by users [15, 50, 58, 30, 59, 60, 61]. According to the findings of the above privacy research, a privacy policy should include: **(1) Data type:** the type of data being collected, **(2) Data access:** who will access the data, **(3) Data usage:** how the data will be used, **(4) Data storage:** where the data will be stored, and **(5) Data retention:** how long the data will be stored. Other studies, such as [62, 63], defined knowledge inference as another factor. Knowledge inference displays what additional information can be learned from the collected data [63]. For example, in a thermostat text privacy notice, individuals are informed that their temperature is being collected to save energy. However, the privacy notice does not include what can be learned from it. Temperature data can reveal a person's location, favorite settings, whether there are visitors in the house and more information that the user is unaware of. In our initial design, we included knowledge inference in PrivacyCube; however, in the first PrivacyCube evaluation (see Section 4.1.1), most participants ignored it and commented that it takes up unnecessary space, so we remove it. Hence, PrivacyCube's final design visualizes five main privacy policy factors: data type, usage, storage, retention, and access.

PrivacyCube further visualizes IoT devices and links them to each privacy policy factor. By integrating the identification of IoT devices and their privacy notices into one design, users can see detailed information about IoT data practices that will enable them to understand the implications of their privacy choices. Based on previous work, we identified the most commonly used IoT devices in each room, and we selected manufacturers based on user reviews [64, 65, 66, 61]. Figure 15 shows the IoT devices selected for each home location. As data practices can vary across devices from different manufacturers, it should be noted that the sensors and data displayed on the cube are only examples; other researchers can use PrivacyCube with a different list of IoT devices to visualize privacy-related notices.Following the devices' selection, we created Figure 1 by generating a list of data types, usage, and access. Figure 1 presents a mapping of the IoT devices displayed on PrivacyCube using multiple privacy policy templates included in the IoT Assistant App [17]. As the IoT Assistant app is a state-of-the-art app that shares a similar focus to our research in its relevancy to IoT and data collection practices, we used its existing templates. We translated its content into data physicalization, rather than recreating or extracting information, providing users with a more user-friendly approach to understanding IoT data privacy. If the IoT device's privacy policy template is not found in the IoT Assistant App [17], or if the template does not cover all five major privacy factors, we manually extract the information from the device's text privacy policy. Table 7 in Appendix B lists the IoT devices we used in PrivacyCube and their privacy policies.

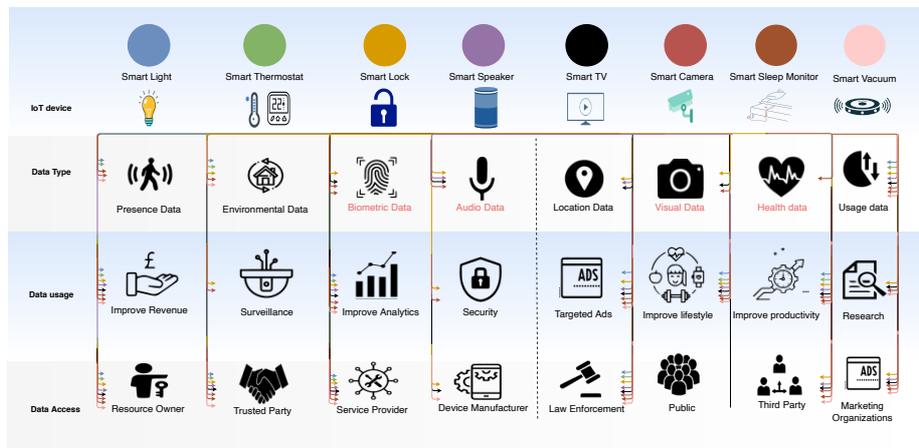

Figure 1: Mapping IoT devices on PrivacyCube to indicate data type, use, and access. Each IoT device has a unique color that is indicated at the top of the device icon. Identifiable data is represented in red, while non-identifiable data is in black.

### 3.2 Usable

PrivacyCube uses a combination of modalities to deliver IoT privacy-related practices As previous research suggests, accompanying icons with text in privacy policies can aid usability and comprehension [16]. Privacy notices are delivered visually through the cube via text, icons, and colored lights to provide users with information that requires minimal to no effort. Drawing from the design principles outlined in the IoT Assistant App [17] and our review of numerous privacy policies (refer to Table 7 in Appendix B), we have carefully designated icons to represent distinct IoT devices, data types, data use, and data access. We accompanied each icon with text for clarification. We added a world map depicting the continent where the IoT data is stored to display data location in the $Lface$. We used LED light to show data collection to help users understand privacy information quickly and accurately. Lighting the LED indicates that the IoT device is collecting data while leaving the LED off indicates otherwise. The cube is sized to be seen from a distance while also taking up no space in people's homes. Users can further explore available IoT device privacy notices via a touch screen at the top of PrivacyCube, eliminating the need for a mobile app or web interface.

### 3.3 User Awareness

One of our primary goals is to simplify learning and encourage privacy engagement, particularly in settings such as homes where people can interact with the device collectively. Unlike standard devices such as hubs [67] designed for individual use and enabling navigation through various applications, PrivacyCube serves as a physical device rather than a traditional screen. This physicality is necessary for creating data physicalization and shared experience that overcomes the limitations of individual screens. Therefore, to avoid overwhelming the user with multiple types of privacy information on one screen, we used each cube face to deliver one (or at most two) pieces of information. Except for the L face, which represents two privacy factors (data location and retention), each of the other cube faces represents one privacy factor. Furthermore, each cube face has a letter label in the top left corner to assist users in identifying the cube faces. The RGB aspects used in the cube also clearly convey to the user the sensitivity of the collected data. This feature can make users more aware of nearby IoT devices and their privacy concerns. We used the three-color approach applied in *Privacy Bird* [3] to define the sensitivity of the collected data, which leverages the well-known traffic light metaphor. Table 1 summarizes PrivacyCube's three-color approach.

Table 1: PrivacyCube uses a three-colour approach via LEDs to indicate the sensitivity of the data it visualises.

| LED colour | Risk level | Representations |
|---|---|---|
| Green | Low risk | Non personally identifiable data |
| Yellow | Medium risk | Neutral data |
| Red | High risk | Personally identifiable data |

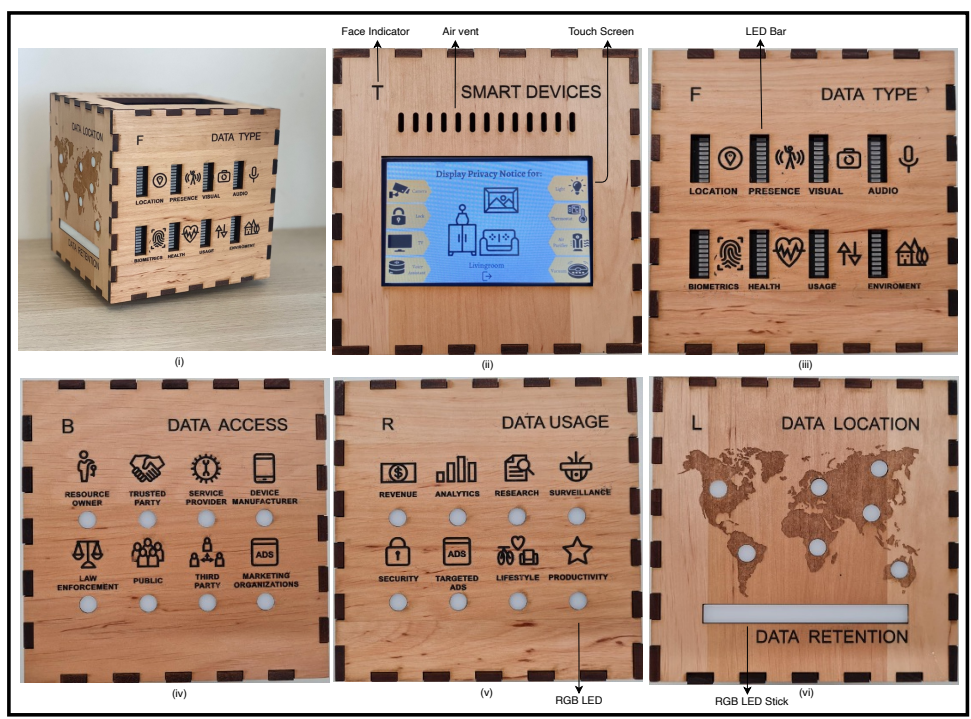

Figure 2: PrivacyCube Prototype (i); PrivacyCube has five faces: (ii) IoT resources, (iii) collected data types, (iv) data access, (v) data usage, and (vi) data storage location and retention period. (Demo Video)

## 3.4 Neutral

Neutrality is defined in privacy research [17, 68, 69] as determining if the privacy interfaces make it difficult for users to exercise their privacy options. Available privacy visualizations frequently nudge users to prefer one option over another. Furthermore, as mentioned in Section 2, the majority of privacy visualizations rely on the use of Web or mobile tools, necessitating more user effort [15, 14]. In this context, we define PrivacyCube as a neutral data physicalization privacy notice that can be deployed in the user's home without needing Web or mobile tools, allowing even non-tech-savvy users to interact with it. We build Privacycube with high-quality wood, similar to most home furniture, to achieve a neutral design. We created the cube to be aesthetically pleasing so that a person can place it in their home without causing disruption. The cube has a rotating base, giving users a more accessible and enjoyable experience interacting with their privacy notices.

## 3.5 PrivacyCube Functionality

This section, presents a fictional use case scenario about Alex to show how PrivacyCube works and how it delivers information to individuals to increase their privacy awareness. Alex protects their home with a smart lock that includes a camera. Alex finds it convenient to know and communicate with visitors to their house via their smart gadgets. Alex does not realize that for the smart lock to work, it must detect and store sensitive data that may be utilized for purposes other than home security. The implementation of PrivacyCube will notify Alex about their data as follows:

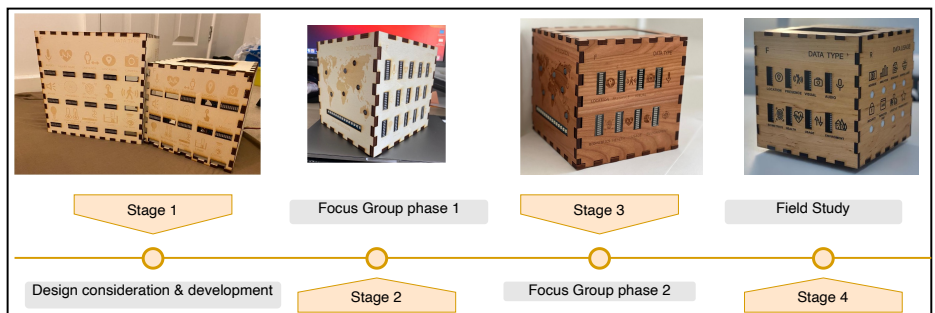

Figure 3: Iterative development of the PrivacyCube interface, with enhancements and testing at each stage. This process is informed by feedback gathered from study participants, which contributes to the refinement and validation of the design.

- The lock icon on the top face will turn green, indicating the IoT device is active.
- The second led for the segment in the location, visual, audio, biometrics, usage, and environment data types icons on the front face will light up, identifying the type of data collected by the second IoT device.
- The resource owner, trusted party, service provider, device manufacturer, law enforcement, third party, and marketing organizations will all light up, indicating data access.
- The revenue, surveillance, analytic, security, targeted ads, lifestyle, productivity, and research icons will light up, indicating data usage.
- The time bar and world map will light up some sections showing where and how long the data is held.

Compared to traditional text privacy notices, Alex now has greater information about the processes carried out on their data.

## 4 Evaluation

Using the four design principles outlined in Section 3, we implemented PrivacyCube to aid users in gaining a better understanding of data privacy. To evaluate the design principles of PrivacyCube, we conducted two evaluations (two phases consisting of a focus group and a field study) to answer the following research questions:

1. Comprehensive: Does PrivacyCube provide users with a sufficient understanding of their IoT privacy data?
2. Usable: Does PrivacyCube provide usable privacy notices?
3. Neutral: Does PrivacyCube motivate users (mainly family members) to have non-biased conversations about IoT devices and privacy on topics such as future purchasing decisions, data control, etc.?
4. User awareness: Can privacy awareness among family be improved by deploying data physicalization that visualizes how IoT devices process data in a smart home environment?

Throughout development, PrivacyCube went through an iterative design process that involved multiple stages of development and testing, including adjustments to icons, text, and several faces in each stage. We updated PrivacyCube based on user feedback after each stage and tested it with participants in the next stage. This allowed for a thorough examination of different design options and ultimately led to the development of an effective interface for mapping IoT devices and properties to a visual language. Figure 3 depicts PrivacyCube's early designs until the last design in stage 4, which we used to conduct the field study.

### 4.1 Study I: FOCUS GROUP

**Participants:** We conducted two phases of a focus group with the same six participants. We advertised on social media to recruit participants who have at least one IoT device in their homes. Doing so will ensure that participants are familiar with IoT. We also asked if they had an information privacy background. Based on this, we carefully selected participants to include individuals with/without an information privacy background to encourage conversations among participants with varying levels of expertise [70, 71]. In a pre-study survey, we asked participants to provide demographics. Of our participants, five identified as female and one as male, aged between 29 to 44 (M = 34.5, SD = 4.89). All participants reported having IT expertise, and 50% of the participants had a background in information privacy. Finally, we asked if they were interested in reading and learning about privacy notices and policies. Five participants stated that they had not read any privacy policies in the previous three months; however, five participants indicated that they read (specifically, skimmed) privacy policies only if they were interested. This finding supports our hypothesis that users want to understand privacy notices but find them difficult to read. The two focus group phases took place in a conference room. We contacted participants via email before the study to schedule the time and to attach the information sheet and consent form. All participants returned the consent form electronically. We conducted the study after receiving ethics approval.

**Procedure:** On the day of the study, we started by explaining the goal of the focus group and outlining the session structure. We divided each focus group session into multiple tasks (described in detail below in 4.1.1 and 4.1.2). Participants were encouraged to share their experiences interacting with privacy policies throughout the session. With the participants' permission, the moderator audio-recorded the session and took occasional pictures to document the study for future analysis. At the beginning of the study, participants were told they were free to leave the study at any time.

During the study, participants interacted with two IoT devices' privacy policies (smart lock [72] and smart sleep monitor [73]). We chose these devices because they collect sensitive data, such as location and health, which increases the viability of informed privacy decision-making. Ring and Jablotron were chosen due to similar policy lengths and positive user feedback.

We asked participants to extract information and respond to the questions (tabulated in Table 4 in Appendix B) via an online survey. We adopted the questions from previous research on privacy policy visualizations [20]. Participants had 20 minutes to extract the information and then 5 minutes to fill out the post-test survey, which included a System Usability Scale (SUS) questionnaire [74] and an adjusted user experience (UE) questionnaire [52]. The SUS is a ten-item questionnaire with Likert-type responses on a scale of 1 to 5 (lower is better for even questions and higher is better for odd questions, 1=strongly disagree, 5=strongly agree) [74, 68]. The UE is a four-item questionnaire on a scale of 1 to 5 (higher is better, 1=strongly disagree, 5=strongly agree). The online survey allowed us to record participants' replies automatically and collect data on how long it took them to extract information. The replies were given on a five-point scale: "Definitely yes," "Probably yes," "Unsure," "Probably no," and "Definitely no." Participants answered questions using a personal device. A moderator observed the participants' attitudes and interactions and gave help as needed.

### 4.1.1 FOCUS GROUP Phase I

**Procedure:** This session lasted 1.5 hours, including a break, and consisted of the following three tasks:

- **Extracting information from a text-based privacy policy:** We provided each participant with a printed copy of the privacy policy for the two IoT devices (smart lock [72] and smart sleep monitor [73]). We instructed participants to answer based on the information included in the printed privacy policies rather than their prior knowledge of the company's privacy practices. Following the first activity, a moderator led a five-minute discussion with the participants regarding their user experience engaging with the text privacy policy representations. The purpose was to gather qualitative feedback from participants and brainstorm solutions for better privacy policy representation.

- **Designing privacy policy solutions:** We presented participants with an empty cube so they could brainstorm solutions. Participants were given sticky notes and asked to draw their thoughts on what constitutes a meaningful privacy notice. The participants were designing solutions reflecting on the same two IoT devices. Figure 4 depicts participants brainstorming solutions for IoT privacy notices. Participants had 15 minutes to complete this task.

- **Extracting information from PrivacyCube:** Following the second activity, we presented the PrivacyCube prototype and explained its role in a smart home environment. We then operated PrivacyCube and asked the participants to extract information from PrivacyCube to answer the online survey form (Table 4 in Appendix B). In this task, participants were instructed to answer based on the information included in PrivacyCube. We were observing participants performing the same activities with and without PrivacyCube, allowing us to assess the usefulness of the data physicalization privacy notices. The study concluded with a five-minute group discussion about the participants' interactions with PrivacyCube.

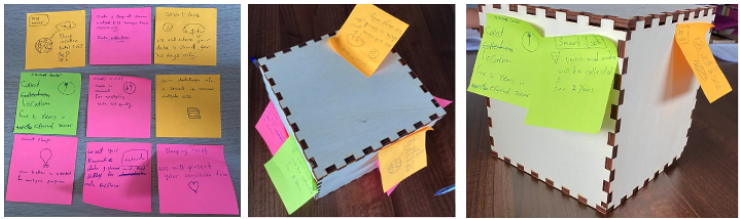

Figure 4: In the first focus group's second task, participants brainstormed solutions for effective IoT privacy notice interfaces using an empty cube and sticky notes.

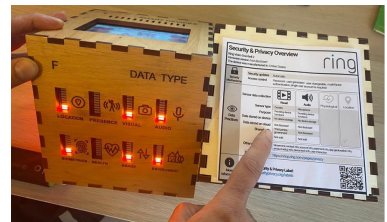

Figure 5: A participant comparing PrivacyCube and Privacy Label in the second focus group's second task.

#### 4.1.2 FOCUS GROUP Phase II

**Procedure:** We modified the PrivacyCube representation based on the findings of the first phase. This focus group session, which we held ten months later, lasted two hours and included the two activities below:

- **Confirming the overall representation of PrivacyCube before testing it in a smart home:** We placed the modified PrivacyCube on the table for the participants to explore. The moderator was observing the participants' reactions as they explored the modified PrivacyCube. We then held a ten-minute discussion during which participants were asked to rate the new PrivacyCube design.

- **Extracting information from Privacy Label:** Following the first task, we conducted a comparison study to compare the PrivacyCube interface to a state-of-the-art interface, the Privacy Label [75]. We created prototype boxes for the same IoT devices and attached primary-layer labels to each. The secondary-layer labels were placed on a mock-up website. The sleep monitor labels were generated using the label generator [76], and we used the *Ring Video Doorbell 2* featured label example for the smart lock labels [77].

  We presented the two prototype boxes and explained the attached labels. Participants were asked to extract information from the corresponding Label to answer the survey form (Table 4 in Appendix B). We instructed the participants to respond based on the information contained in the labels. Following the activity, we conducted a 40-minute semi-structured discussion to learn how the labels compare to PrivacyCube [78]. The discussion questions are tabulated in Table 6 in Appendix B. Figure 5 shows a participant comparing PrivacyCube to Privacy Label.

**Focus Group Phase I and II Quantitative Results** Overall, the quantitative results of the focus group showed that PrivacyCube receives a better rating compared to both text and label privacy policies. We first ran the Shapiro-Wilks Test [79] and the D'Agostino's and Pearson's Test [80] and confirmed that the participants' replies were normal. We then performed a One-way repeated measures Analysis of Variance (ANOVA) to analyse the participants' responses. The one-way ANOVA shows that PrivacyCube outperformed the text representation in all dimensions, with PrivacyCube outperforming the label representation specifically in terms of usability (F (1.366, 6.831) = 207.2, P<0.0001). Figure 7 depicts PrivacyCube's SUS score in comparison to text and label. PrivacyCube's mean SUS score across all participants was 80, which is above 68 points, meaning PrivacyCube is usable. We present the results of the first two research questions below:

- *RQ1: Comprehensive.* We analyzed the perceived difficulty of PrivacyCube and compared it to text and label representations measured using items 5, 6, 7, and 10 from SUS [74, 68], and item 2 from the UE survey. The results show that the average PrivacyCube score is higher in the four SUS items related to difficulty, i.e., higher in items 5 and 7 and lower in items 6 and 10. Participants also felt more informed about how smart devices use data after extracting information from PrivacyCube, which is confirmed through item 2 from the UE survey (for full SUS/UE items, refer to Figure 16 in Appendix B.) Overall, participants rated both PrivacyCube and the labels as comprehensive and understandable because they contain the essential information users need to know about their privacy.

- *RQ2: Usable.* We used multiple quantitative measures to analyze the usability of PrivacyCube. We used the information extraction tasks to compare participants' ability and effort through the following metrics [68]:

  **Accuracy and Confidence.** As a measure of user accuracy, we counted the number of participants who correctly extracted information and answered the questions during the information extraction tasks. Participants' answers were deemed accurate if they chose "Definitely yes" or "Probably yes" when the correct answer was yes, and "Definitely no" or "Probably no" when the correct answer was no. As a measure of user confidence, we counted the number of participants who answered every question both confidently and accurately (for instance, answering "Definitely yes" when the correct answer was yes). Figure 6 depicts the mean rates of accuracy and confidence. We used a One-way repeated measures ANOVA to analyze our results on accuracy and confidence. The results were statistically significant for confidence (F(1.385, 12.47) = 7.546, P<0.01) and not significant for accuracy (F(1.425, 12.83) = 3.409, P>0.05 ). Pairwise Post-hoc comparisons using the Tukey HSD test indicate that participants demonstrated improved accuracy and confidence using PrivacyCube compared to the text privacy policy (for accuracy, M= 3.6, SD= .741, p= 0.0458; for confidence, M= 1.95, SD =.606, p= 0.0010). On the other hand, we found that there were no statistically significant differences between the label and PrivacyCube in terms of accuracy and confidence (for accuracy, M= 4.3, SD= .025, p= 0.9703; for confidence, M= 2.7, SD= .0637, p= 0.8171).

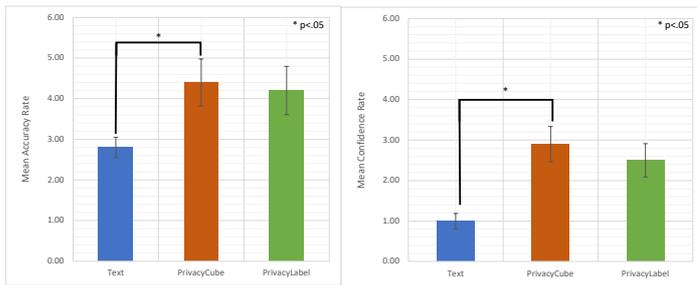

Figure 6: Mean rates of accuracy and confidence on the two information extraction tasks, shown with standard error bars.

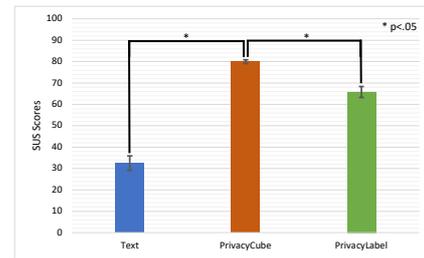

Figure 7: System usability scale (SUS) scores for text, PrivacyCube, and Privacy Label representations.

**Average time taken to complete the extraction tasks.** Participants spent more time completing the extraction tasks using PrivacyCube and privacy labels: an average of 12 minutes for PrivacyCube, 11.16 minutes for labels, and 8.5 minutes for text. This was expected as we observed all participants answered the questions randomly and did not read the text representation privacy policies. On the other hand, we noticed that all participants were easily following the labels and attempting to extract information. Four participants also opened the secondary layer. With PrivacyCube, participants were socially active and excited to find and discuss the information, which explains why they took longer. Hence, participants' answers were more accurate using PrivacyCube and labels, as described above.

**Perceived effort.** We compared PrivacyCube's perceived effort to text and label policy representations using items 2, 3, 4, and 8 from SUS [74, 68], and items 1, 3, and 4 from UE survey. The results show that the average PrivacyCube score is better in the four SUS items relating to the ease and difficulty of participants' experiences, i.e., higher in item 3 and lower in items 2, 4, and 8. Compared to the text and label representation, participants felt more comfortable and enjoyed using PrivacyCube (items 1 and 4 from UE survey). Participants also expressed that they are more likely to use PrivacyCube as a privacy policy representation (item 3). For full SUS/UE items, refer to Figure 16 in Appendix B. Overall, participants described both PrivacyCube and labels as simple tools that require minimal effort.

**Focus Group Phase I and II Qualitative Results** We transcribed the sessions' recordings and combined them with the moderator's notes, the moderator's occasional pictures, and the participants' notes from the study tasks. The data provided extensive qualitative insights into individuals' attitudes toward the three representations and emphasized the importance of design principles (i.e., comprehensiveness, usability, neutrality, and user awareness). We identified five main themes following [81]: Clarity of Information and Comprehension, Engagement and Interest, Privacy Awareness, Interaction/Communication Facilitation, and Content and Visual Design. The qualitative findings of the post-extraction task discussion revealed that participants had difficulty extracting information from text privacy policies. All participants reported that reading is not difficult, but presenting a large amount of information makes it difficult for the user to read. However, participants reported that extracting information from both PrivacyCube and labels was easy to understand and enjoyable.

- **Clarity of Information and Comprehension.** During the first phase, P1, P2, and P4 expressed confusion and a lack of understanding about the benefits of text privacy policies. One participant (P5) acknowledged the legal requirement to have text policies but agreed that they are difficult to comprehend. On the contrary, all participants found the PrivacyCube visualization effective in drawing attention to their data and improving their understanding of privacy. P1, P2, and P4 found PrivacyCube to contain all the necessary information without the need for extensive reading. The other participants appreciated PrivacyCube's conciseness but felt that it could be improved by providing information about data sharing. In addition, P3 and P6 found the "inference knowledge" aspect hard to understand and suggested a change to how this information is presented, which we did in the second phase.

In the secondary phase, all participants agreed that PrivacyCube and the label provided the necessary privacy information. All participants found the label design intuitive and confirmed that PrivacyCube could be more intuitive and straightforward once explained. The label's multi-layer feature was well received by the participants, as they noted that it provides more elaboration. Nevertheless, PrivacyCube was still preferred as it was simpler, easier to understand, and provided information faster. *"I see the cube a lot better. It allows me to find information faster. It only shows related lights, making it easier and faster to understand. The label is also easy, but I have to do an extra reading; with the cube, all I need is a spin without reading from top to bottom,"* commented P3. Participants did not find the label difficult to understand but commented that it required more effort to read and comprehend compared to PrivacyCube. Participants also noted that the icons and light colors in PrivacyCube helped them comprehend the data, as it draws attention quickly to the used data. One participant (P6) mentioned the advantage of clarity and specificity in the labels, particularly for data retention periods. *"The label was only clearer in the retention, Privacycube could tell you that your data is collected instantly [...], but if you want the exact number, the label is more specific."*

- **Engagement and Interest.** In the first focus group, the moderator noticed that no participants were interested in extracting information from the printed privacy policies. As expected, all participants stopped reading the printed privacy policies and answered the questions randomly after a few minutes. The three participants with no previous privacy background commented that they would not read any text privacy policies. Participants with privacy knowledge reported that they are willing to read only if they think the IoT device collects sensitive data. On the other hand, participants were engaged while extracting information from PrivacyCube. Participants stated that PrivacyCube would be a great addition to their homes as it is fun and as it presents every IoT device. P3 commented, *"I liked the data storage face, so much fun looking at the map"*.

  Throughout the second focus group, the moderator observed that the participants were following the label to answer the questions, with four participants accessing the secondary layer. When asked about their interest and potential long-term usage of the label and PrivacyCube, all participants agreed that PrivacyCube was more exciting and intriguing than the label. The participants considered the PrivacyCube's physical nature made it more attractive and engaging. In addition, the interactive visual display of PrivacyCube was deemed to have aided in presenting information in a more understandable and accessible way. As one participant noted *"It's not reasonable that every time I will check the label or open a website to read the information [...] the cube is so concise, fast, quick in picking out the information."* Furthermore, participants remarked that PrivacyCube featured fewer texts than the label, resulting in a more aesthetically appealing product.

- **Interaction/Communication Facilitation:** At the first session, participants were immediately intrigued by PrivacyCube and began exploring its icons and lights. All participants were involved in discussions surrounding privacy, with several discussing their understanding of data collection. Participants found PrivacyCube a convenient and straightforward alternative to text policies. Some scripts of participants' discussion, *"The light on/off feature is quite convenient," "It is simple and straightforward, unlike text."* Participants confirmed that PrivacyCube effectively communicates privacy notices and expressed eagerness to learn more about IoT privacy using it. P4 suggested that PrivacyCube could be used in stores to provide information about new devices, with P2, P3, and P5 expressing interest in having the cube in their homes for added peace of mind. Participants also stated that by using PrivacyCube, they could teach their families how IoT work in an easy and understandable manner.

  During the second session, participants discussed the effectiveness of communicating privacy information through different mediums. The label, although appreciated, was seen as functional but lacking in innovative features compared to PrivacyCube. Most participants commented on the label's advantage regarding its ability to be opened on different devices and how easily it provides access to extra information. P5 noted that they prefer if there was an electronic version to complement PrivacyCube, similar to the label, so it can be accessed using multiple devices. (P3-6) also noted that PrivacyCube should have the feature of allowing for additional information access. Nevertheless, all participants still commented that PrivacyCube's interactive features and engaging design provided an easier way of communicating privacy information. As an example, P3 commented, *"I find the cube easier to use and explain. It's something that people can see in front of them [...] even in any lecture, if the instructor gives us a paper, a physical device, and something on the phone, I believe most would be inclined to learn using the device. This is why I think that using the cube is better for teaching privacy because the first step in teaching is grabbing people's attention, and the label will not do that."* As with the first phase, all participants were engaged in talking about privacy around PrivacyCube and noted that its visual design and aesthetics allow for more conversations around privacy. Participants also discussed the accessibility and user-friendliness of the label and PrivacyCube for different age groups and individuals with disabilities; refer to Table 2.

- **Privacy Awareness:** In the first focus group meeting, P2, P3, P5, and P6 reported feeling hopeless when obtaining an IoT device because text privacy policies fall short of providing privacy awareness. The participants acknowledged that they do not realise that small IoT devices can pose privacy risks. For example, P4 and P1 expressed concerns after learning that the smart sleep monitor could sell their health data to insurance providers. Commenting on PrivacyCube, four Participants (P2-5) reported that having PrivacyCube is essential for raising their awareness of IoT privacy. Specifically, three participants (P1, P2, P4) highlighted PrivacyCube's effectiveness in informing them of IoT data collection and its diverse applications. All participants indicated that a physical device capable of continuously alerting them to their IoT devices' data collection is crucial in their homes. P1. said *"It is a beautiful idea, many people need it, I need it myself, my mum needs it, and my children need it."*

  At the second focus group meeting, we noticed that PrivacyCube and the label have different impacts on privacy awareness. All participants reported that PrivacyCube provides a more visual and physical format that enhances accessibility and awareness of data privacy practices. In contrast, they valued the detailed information and linked resources the label provides. As P6 stated, *"both increased my awareness. While I prefer the cube ease of showing information, I have a better understanding through using the label."* Additionally, P1-4 noted that while the label may be more familiar to some users, PrivacyCube provides an immediate visual representation of privacy practices that users can easily grasp. P1, P3, P5, and P6 indicated that the feedback provided by the cube through its lights was helpful in quickly identifying which devices were collecting data and the potential risks associated with such data collection. In comparison, they noted that the label required more effort to read and understand.

- **Content and Visual Design:** Throughout the first phase, all members participated in developing more effective and usable privacy solutions, see Figure 4. Participants mainly recommended using fewer words, using bullet points, using well-known icons, and including only sensitive information, such as data collection, retention, and sharing. Study participants liked using the empty cube and stuck their notes on the side closest to them.

  Group members discussed that the design of text privacy policies provides an unusable solution. Except for P1, all participants praised PrivacyCube's design and liked the shape and layout. P1 stated, *"I would prefer if it was designed like a picture frame [...], it is a little difficult to rotate."* However, P4 stated, *"It is such a nice design, and the cube gives it a unique look."* P5 suggested adding a rotating feature to the cube's bottom to make it easier to interact with. All participants suggested using high-quality wood and reducing the cube size while increasing the screen size to boost interactivity. Participants appreciated that each cube face represents one data aspect, which they found useful, simple to understand, and easy to remember.

  In the second stage, we noticed that participants were impressed by the new design. They all commented on how beautiful it is and how it can deliver privacy information in a fun interactive way. All participants liked the wood material and the addition of the *lazy Susan*. Participants also agreed that the new cube's size and the new touchscreen complement each other. Participants confirmed that the cube has a product-like design and that they would like to buy it one day. P2, P4, and P5 suggested painting the icons for better visibility, while P1 and P6 suggested dimming the lights to make them more comfortable. Overall, participants liked the new design and said it was ready to be included in any smart home.

  After that, participants shared their thoughts on the label's and PrivacyCube's content and design. All participants appreciated the label's two-layer structure, which allowed them to select the level of detail they desired. They noted that the label was well-organized and easily accessible for detailed information, whereas PrivacyCube offered a more visual representation for general use. However, participants thought the label was not very interactive, with P1, P4, and P6 comparing it to an electronic paper, card, or PDF. *"The label only changes privacy policies from long text to brief text with figures; there is nothing new. It wasn't exciting; it simply does the job because it is organized,"* P5 commented. Some participants mentioned that the label had some drawbacks, such as the small text (P4) and the need for an additional device, e.g., a phone, to access the secondary layer (P6). P1, on the other hand, appreciated the ability to access the label via their phone, allowing them to zoom in and out for better legibility. In comparison, PrivacyCube was found to be more interactive and easier to use. All participants appreciated its use of light to indicate the status of data collection and its spinning feature for accessing information. P1 suggested that PrivacyCube and the label could complement each other, whereas P2 and P4 believed that if PrivacyCube included a feature to access detailed information by clicking on the screen, the label would be unnecessary. P3 also stated,*"The cube is more user-friendly, [...] the information comes to me embedded in an interactive way. The labels' icons are not physical or eye-catching, there is no colour or movement in them."*

Table 2: Evaluation of privacy policy interfaces: Text, Privacy Label, and PrivacyCube. Findings from the Focus Group Study.

| Aspect | Text | Privacy Label | PrivacyCube |
|---|---|---|---|
| Interactivity | Not interactive | Less interactive | More interactive |
| Amount of Info | Heavy text | • Less text-heavy<br>• Includes detailed information and links for more information | • Light text<br>• More visual with icons and color-coding |
| Accessibility | Difficult to locate | Can be accessed through multiple devices | Requires possession of the tool |
| Intuitiveness | N/A | Intuitive | More intuitive once understood |
| Engagement | Not engaging to use | Less engaging to use | More engaging to use |
| Recommended Users | • Recommended for legal entities<br>• Not meaningful for the consumer | • Recommended for those interested in privacy<br>• May not be suitable for those with Dyslexia | • Suitable for people of different ages<br>• May not be suitable for those with color blindness |
| Technical Jargon | Overly technical | Somehow technical | Simple |
| Ease of use | Requires heavy reading | Requires some reading to access all information | Concise and quick to pick out information |
| Design | Long and complex Text-based | • Can have an adjustable size<br>• Can provide zooming<br>• Lacks colour usage | • Static size<br>• No zooming<br>• Traffic light system |

## 4.2 Study II: FIELD STUDY

Based on the findings of the focus group, we built two PrivacyCubes for use in the field study. We conducted the field study with eight households, i.e., two homes at a time. This method helped us to troubleshoot any technical issues faced by participants during the trial.

**Participants:** We recruited eight households by advertising on the university mailing list. In the email, we mentioned the qualification criteria as households with at least two members living together and at least two IoT devices in their homes. Our goal was to target families with basic IoT knowledge to encourage learning about IoT privacy. The participating households included six families with children and two families with only adults; refer to Table 3. We gave each household a £50 shopping voucher to appreciate their participation.

**Procedure:** We conducted a two-week field study consisting of 8 semi-structured interviews with one member of each family. We borrowed from technology probe [82] and deployed PrivacyCube in real-world settings to encourage individuals to make informed privacy decisions in line with their habitual IoT usages. Although we borrow from a technology probe, our in-home deployment of PrivacyCube does not log data. We preferred direct engagement with participants because the primary goal of PrivacyCube is to encourage learning and exploration of privacy. We relied on the PrivacyCube diary and interviews to collect feedback.

We initially met with the participants in the university's smart home lab. We started by outlining the study's goals and ethics. We then asked the participants to sign the consent form and complete an online awareness survey to measure their IoT privacy awareness before using PrivacyCube. The awareness survey includes questions about four IoT devices' privacy practices (Table 10 in Appendix B). Following that, we showed the participants PrivacyCube, explained its role, and explained what each face of PrivacyCube means. We also handed the participants a PrivacyCube instruction sheet with a written explanation. PrivacyCube was then demonstrated in the smart home lab using two smart devices: a smart speaker and a smart vacuum. Participants were encouraged to ask questions during and after the demonstration. We explained to the participants that PrivacyCube would light up some LEDs in their homes at regular intervals to simulate the capabilities of IoT devices.

Table 3: Demographics of PrivacyCube's field study participants.

| # | Members | IoT devices | Privacy Background | SUS |
|---|---|---|---|---|
| h1 | All male aged (25, 27, 30, 31) | Smart switches & Smart TV | One member | 90 |
| h2 | M:45, F:38 with four children (f:16, f:7, m:5, m:1) | Smart speaker and Smart TV | None | 77.5 |
| h3 | All male aged (30, 35) | Smart Thermostat, Smart speaker, & Smart light | None | 90 |
| h4 | M:36, F:32 with three children (m:7, m:5, m:1) | Smart Camera & Smart TV | None | 72.5 |
| h5 | M:47, F:44 with three children (f:17, f:6, m:5) | Smart lock & Smart TV | None | 87.5 |
| h6 | M:39, F:33 with three children (f:10, f:7, m:4) | Smart Thermostat, Smart vacuum, Smart light & Smart TV | One member | 92.5 |
| h7 | M:35, F:32 with two children (f:6, f:1) | Smart Thermostat, Smart Camera & Smart TV | One member | 80 |
| h8 | M:38, F:30 with two children (f:7, f:3) | Smart speaker & Smart TV | None | 100 |

We asked the participants to move PrivacyCube between rooms in their homes over the two weeks: four days in the living room, four days in the kitchen, three days in the bathroom, and three days in the bedroom. To encourage privacy conversations, we let the participants choose the order of the rooms based on where the family usually gathers. Lastly, we handed out the diary sheets explaining that one family member had to fill out a diary for each room. We also sent the participants an electronic version of the diary sheets where they could choose what works best for them. Table 5 in Appendix B contains the diary questions adopted from [83]. One week through the study, we sent a follow-up email to each family, thanking them again for their participation and reminding them to move PrivacyCube between rooms and to complete the PrivacyCube diary.

At the end of the two weeks, we emailed the participants asking them to fill out a SUS questionnaire and the same online awareness survey they completed at the first meeting. This will allow us to test PrivacyCube's usability in a home setting and assess participants' IoT privacy awareness after using PrivacyCube. We also scheduled an interview with a member of each family in an agreed-upon location. At the interview, We began by collecting PrivacyCube and the diary sheets. We reviewed the diary sheets and asked for clarification on each annotation. Following that, we asked the participants about their use and deployment of PrivacyCube in their homes. Table 8 in Appendix B contains a complete list of the questions. At the end of the interview, we thanked the participants, gave them our contact information, and asked them to send us any comments they wanted to share. With the participants' permission, the interview was audio recorded.

To build the IoT awareness survey, we borrowed from [84, 64] and adjusted the questions to situate PrivacyCube. In the survey, we asked the participants to fill in 20 questions to evaluate their privacy awareness level of IoT devices. The questions were multiple-choice (multiple answers), distributed as five multiple-choice questions per simulated IoT. Each question has a note that briefly defines the IoT device and the multiple choices.

**FILED STUDY Quantitative Results** The quantitative findings demonstrated that PrivacyCube had a positive impact on the home environment. We present the results to the second and fourth research questions below:

- *RQ2: Usable.* Using the post-study SUS survey, we analysed PrivacyCube usability in a home environment on a scale of 1 to 5. The SUS score for all participants (see Table 3) was M = 85.83; SD = 8.96; Mdn = 88.75, which is higher than the 68 points, indicating excellent usability, according to Bangor et al. [85]. The phrase "I found the various functions in this PrivacyCube were well integrated" received the highest rating (M= 4.59; SD=.517). The phrase "I think that I would need the support of a technical person to be able to use this PrivacyCube" received the lowest rating (M= 1.09; SD=.353). Five households rated PrivacyCube excellent, with one giving it a perfect SUS score of 100.

- *RQ4: User Awareness.* As a measure of user awareness, we counted the correct answers to the awareness survey that participants completed twice, once before and once after the two-week field study. Each correct answer adds one point to the total, calculated out of 100. Figure 8 depicts the mean awareness level before and after using PrivacyCube. To analyze our results on user awareness, we used the paired t-test. The findings show that participating households demonstrated improved user awareness after using PrivacyCube for two weeks. The results were statistically significant at $p < .05$ (p=0.00041, t= -5.57).

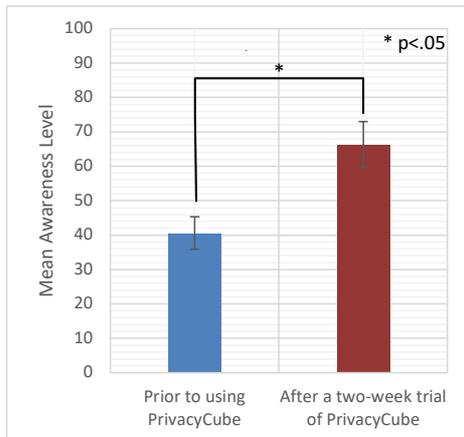

Figure 8: Households' mean awareness level before and after using PrivacyCube, shown with standard error bars.

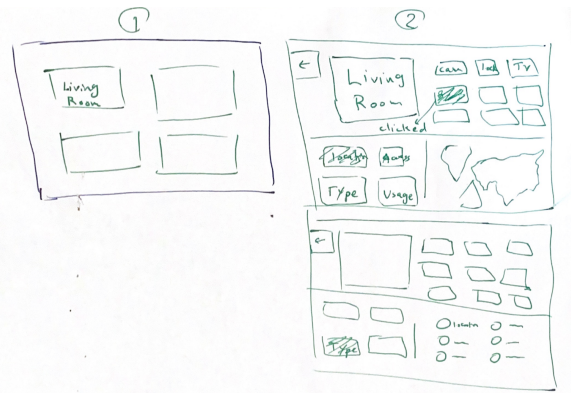

Figure 9: A member of Household 7 created a flat physical privacy interface design by replicating the content of PrivacyCube.

**FILED STUDY Qualitative Results** The results reported here are primarily based on the final interviews. Except for one household that filled out a hard copy of the diary, all other households filled out the diary electronically. These diaries included many details and pictures, including several short videos showing PrivacyCube, family members using PrivacyCube, and overall details of how PrivacyCube is used in a typical home environment. The photos in the diaries allowed us to use the photo-elicitation approach [86] during the interviews, which gave us helpful feedback.

We transcribed the interview recordings and combined them with participants' diaries, photos, and videos. We discussed the findings and performed a thematic analysis in accordance with previous guidelines [81]. The results addressed the third research question (i.e., RQ3: Neutrality) whilst also emphasizing the other research questions. In the following, we present themes that emerged from participants' explanations of their experiences deploying PrivacyCube at home.

- **RQ3: Neutral.** A key aim of PrivacyCube is to provide a neutral privacy interface that enhances home aesthetics and comfort while maintaining all-day privacy awareness and confidence. We sought to create a privacy interface with the best likelihood of blending into the background over time without changing behavior or interpersonal interaction.

  We used a variety of criteria to assess neutrality. First, we assessed participants' acceptance of a privacy interface as a physical device in their homes. All participants mentioned that PrivacyCube was a beautiful decorative piece in their homes. Participants confirmed that the cube shape and material complement the home's overall appearance without causing any distribution. For example, h5 commented, *"it did not look odd; it matched the furniture, I did not feel like I have to tidy it up or put it away."* H1, h2, h4, h5, and h6 all stated that they would like to have a privacy interface like this in their home. H6 stated, *"it is like a vintage product that made my home beautiful."*

  Second, we assessed how well PrivacyCube provided users with a neutral ground to interact with their privacy notifications. Except for h2 and one member of h7, all other households appreciated the cube shape for easier interaction with their notifications. H2 and the adult male in h7 preferred interacting with their privacy notification through a flat physical interface. For h2, their preference was related to the size. For h7, the adult male preference was related to ease of use, as described by the adult female in h7 *"my husband designed an entire design. He said the cube is creative, but whenever he needed to know about data, he needed to turn the device. So he prefers to have it as only one screen."* On the other hand, all other households commented on the ease of use provided by the cube visualization and the fact each cube face represents one privacy aspect. For example, h3 commented *"The cube shape is a very professional idea, through looking at one face, for example, the map I know it is about location, there is nothing else that distracts me, unlike flat designs."* Further, all participants emphasized the cube's rotational feature. For example, h4 also commented by saying *"If it does*

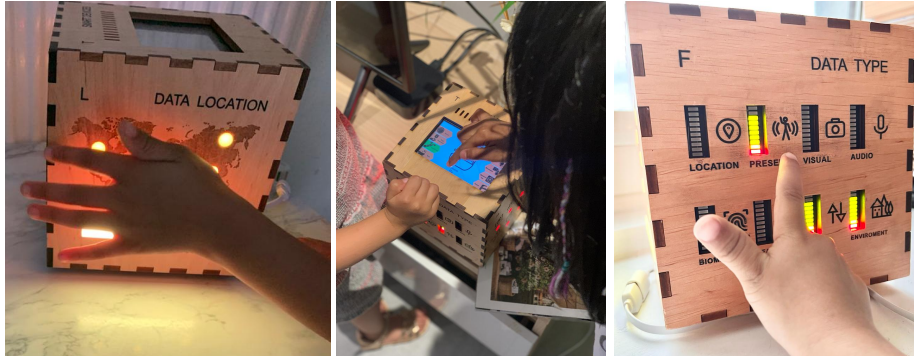

Figure 10: Observations depict participants' children interacting and exploring PrivacyCube, (i) rotating the cube, (ii) tapping the screen, (iii) demonstrating awareness by noting and interacting with on-screen icons.

*not have a movement feature, I will not move it."* This shows that the cube rotation allowed the households to maintain their natural behavior while interacting with privacy notices, highlighting its neutrality in avoiding unintended societal consequences.

In addition, PrivacyCube immediately captured all household members' attention, including children, who began looking at the icons, rotating the cube, and tapping on the touch screen to explore the lights (Figure 10). All families with children mentioned that their children enjoyed exploring PrivacyCube by moving it around the house and challenging themselves with what each smart device collects. Three families shared photos of their children exploring the cube (see Figure 10), and two shared videos of how their children spent time learning by playing with PrivacyCube. H8 said, *"My kids challenged themselves. The oldest daughter asked her brother, "What do you think the TV can collect?".* H7 also commented by saying that the cube was easy to use for the kids. The preceding supports that PrivacyCube provides an accessible and engaging experience for all home users to interact with their privacy notices.

Lastly, we reviewed the diaries and interview responses to determine if PrivacyCube encouraged participants to have non-biased conversations. All households discussed the significance of having PriavcyCube in their homes, with half stating that it would not prevent them from purchasing IoT devices; instead, they would buy devices based on their capabilities and convenience. This echos that although participants liked the appearance of PrivacyCube, it did not affect their natural privacy behavior or nudge them into specific privacy choices. Families also reported that PrivacyCube inspired them to discuss their IoT data in ways they had never considered before. Some data control measures mentioned by participants: turning off the device when not in use (h1, h4, h7, and h8), configuring privacy preferences (h2, h4, h5, and h8), disposing of the device (h3), and looking up information (h1, and h5).

- **Communication.** All families reported that PrivacyCube was most effective when placed in the living room. When we asked the participants about PrivacyCube in the kitchen, the majority commented that it looked neat but preferred if it was waterproof. Two households (h3 and h5) suggested adding sound to PrivacyCube to increase attention in the kitchen because lights can go unnoticed due to having multiple appliances and the focus on meal preparation. Whilst h3, h7, and h8 mentioned that they talked with their families about some unexpected data collection in the bathroom, such as the toothbrush and the bath, all other households agreed that data physicalization privacy notices in the bathroom are unnecessary. Four households did not prefer PrivacyCube in the bedroom due to light brightness (h1 and h4) or little time spent in the bedroom (h2 and h6). On the other hand, other households appreciated how PrivacyCube created unexpected moments of socialization and conversation in the bedroom. H5, h4, and h8 mentioned that bedtime was a good time to discuss what they had learned about IoT privacy with their children or spouses. For instance, h5 commented *"It didn't bother me at all. It gave us some excitement and raised many questions that I found very rewarding to answer as a mum teaching her kids about such a complex topic as privacy."* Figure 11 depicts PrivacyCube in various rooms throughout the participants' homes.

  Every household mentioned that whenever PrivacyCube emits a light, all household members try to guess what it is and where the data goes. Participants also noted that PrivacyCube in the living room sparked IoT privacy discussions within and outside the family circle. PrivacyCube's physical feature and design prompted home visitors to ask questions and interact with the cube's privacy notification.

- **Learnability and Comprehension.** As a critical research objective, we sought to understand how people can learn and engage with their IoT privacy through data physicalization. We focused on these aspects during the interview. Participants noted various ways of interacting with PrivacyCube, which positively impacted their privacy knowledge when using IoT devices. Most households agreed that data physicalization encouraged learning and curiosity. As one interviewee put it: *"Since I bought an IoT device, I honestly have never thought about privacy in my life. [..] You made privacy physical, which is more accessible and easy to understand and think about."* This confirms that participants viewed data physicalization as an effective learning tool. An interesting finding from the interviews was that participants noticed they were thinking about their privacy more frequently and discussing it with friends and family. *"if you have a device like this in your home, and it is showing you the process done on your data, day after day, I will develop more awareness of my data, and I will feel its importance. I even tell my kids and friends to be careful with their data,"* commented h7.

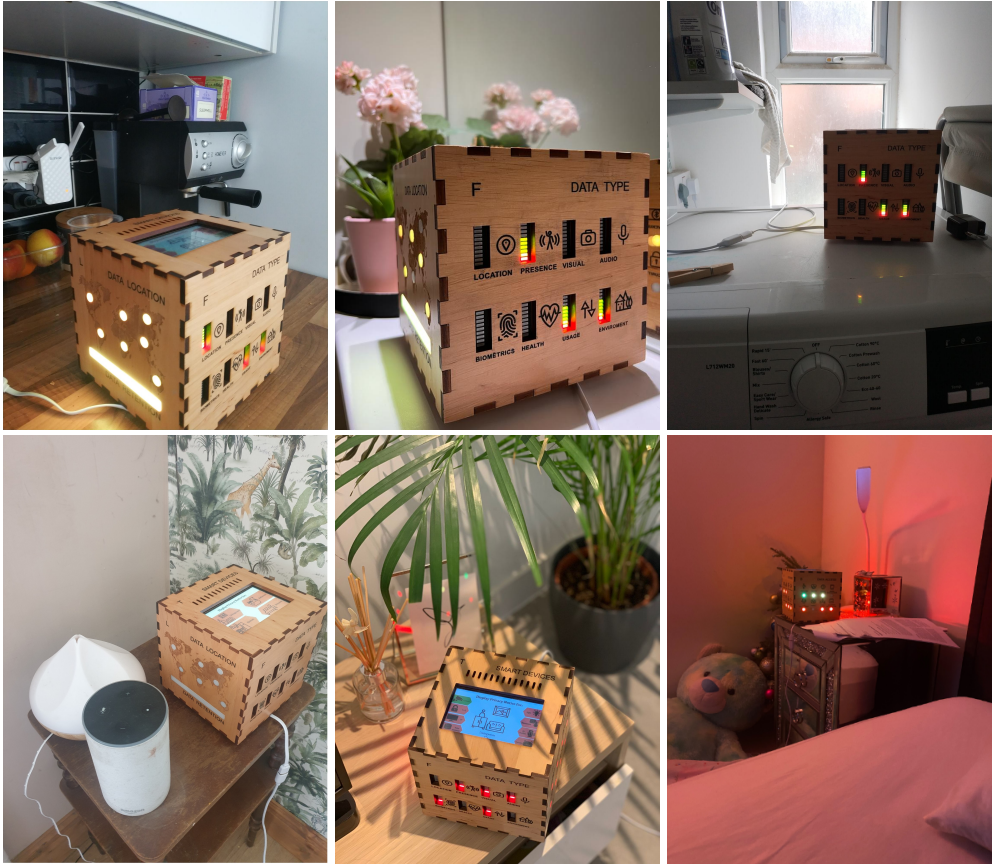

Figure 11: A demonstration of PrivacyCube deployment across the participants' homes showing how it can be used and implemented in different settings.

Participants also reported becoming more aware of the processes performed on their data and a desire to learn more about IoT privacy. H4 said *"When I compare my privacy knowledge before and after using PrivacyCube, of course, I am more aware now."* Several participants noted that they developed learning strategies after a specific period of time. A common view amongst families with children was that PrivacyCube serves as a kid-friendly educational tool with a high teaching potential. As h6 explained *"right after I showed it to my kids, they said, ("oh is this online privacy, we learn this in school.") [..] They do learn privacy in school. Still, all they learn is to keep their password safe. I believe this tool is just as important in the home as in schools for teaching the younger generation about emerging IoT privacy. And the fact that it is in the form of a cube would make teaching much simpler"* Further, those with children reported how attached their children are to PrivacyCube.

When we asked about PrivacyCube's comprehensiveness, the majority stated that it contains everything they are concerned about. For instance, one family noted *"we didn't need to look up more information because that is what we were interested in."* Five participants (h1, h2, h5, h6, and h7) primarily focused on data location and type. Four participants (h5, h6, h4, and h3) stated the importance of data usage. Two participants (h2 and h8) emphasized data retention, while two others (h4 and h5) emphasized data sharing. A small number of those interviewed (h5 and h7) suggested incorporating the guideline paper into PrivacyCube to make it more comprehensive. H5 and h4 wished that PrivacyCube had a feature that allowed them to configure their privacy preferences rather than logging into the preferences of their devices. In addition, while we clearly informed the participants that PrivacyCube sends simulated notices, two households (h2 and h6) expressed confusion only at the beginning of the study.

- **Emotion and Privacy Attitudes** Participants had a variety of emotions in response to the data physicalization privacy notices. The majority of households expressed a high level of excitement during the first five days, but this level tended to decrease as time passed. Five participants (h1, h3, h5, h6, and h7) mentioned that excitement is a recurring emotion; it reappears whenever they plan to bring a new IoT device into their home or whenever a visitor comes to explore the cube. H1 and h6 stated that, while they valued the PrivacyCube notifications, the reappearing light sometimes made them feel scared in their homes. They suggested that PrivacyCube might work well in offices or shared spaces.

    Confidence was also one of the emotions that most participants in the interview mentioned. H5 said, *"now, after using the cube, I have something to back up my arguments. I know what data the devices collect."* Other participants noted their increased confidence after using the cube and their surprise when they realized how their data was used. As h6 stated, *"Yes, I am both confident and surprised. I was shocked to learn they were using my environmental data to generate revenue [..] then I remembered receiving an email telling me that I could only put out three trash bags at a time or the surveillance camera would detect me and impose a fine!"*.

    All participants felt an impact on their privacy attitude after using PrivacyCube. Participants stated that they value their privacy decisions more now that they are aware of how their data is processed. Participants also stated thinking more about and digging deeper into their privacy policies to understand data practices better. Households confirmed that the PriavcyCube made them reconsider purchasing IoT devices and weigh the benefits versus privacy. For instance, h4 said, *"my husband, who never thought about privacy, mentioned that it is better for the person not to buy IoT devices, or a person needs to do privacy configuration himself without relying on anyone and without relying on the default configurations"* Furthermore, participants reported that using PrivacyCube had an impact on their behavior. For example, h4, h5, and h8 stated that they would leave the room to continue their private conversation if they noticed the smart speaker was active.

- **Data Physicalization Design** All participants expressed the benefits of the data physicalization. As mentioned in one of the themes above, h2 and h7 suggested displaying the same cube's privacy notifications on a flat surface to make them more visible and faster to notice. All other families confirmed that the cube shape was easy to use, a great way of presenting information, a great conversation starter, and provides easy access to privacy policies. Several participants also highlighted the cube faces by expressing their preference for accessing privacy policies one piece at a time rather than seeing everything at once.

    Of the eight households who participated in the study, half (h1, h4, h6, and h7) proposed replacing the screen position to make it visible from a distance. They indicated that approaching the cube to check the IoT devices was sometimes too much effort. H6 suggested that the cube's top face could pop out or include a wireless charger to encourage people to interact with the screen. While all participants liked the bars on the data type face, over half said it was sometimes confusing to know which device was on. Some families also suggested fancy PrivacyCube enhancements, such as music support (h6), a button to turn IoT devices on or off (h1), and displaying each IoT device's energy consumption (h4).

## 5 Discussion

This section discusses the benefits of using PrivacyCube for displaying privacy notifications, the limitations of PrivacyCube, and other opportunities for improving data privacy awareness.

### 5.1 The Impact of a Data Physicalization

As described in section 2, effective privacy visualization can enhance users' awareness about data collection and use. Studies referenced in 2 also indicated that data physicalization fosters individuals' engagement and learning. Our work demonstrates that data physicalization can provide users with usable privacy notices that help them understand and communicate about their IoT data privacy.

There are several reasons why data physicalization can help address the lack of IoT privacy awareness in smart homes. Since data physicalizations are usually within individuals' sight, users can quickly notice and interpret privacy notifications without needing Web or mobile tools, which saves time and effort. Privacy data physicalization can serve as home decor, which helps families socialize and discusses IoT privacy. We carefully designed PrivacyCube to communicate comprehensive privacy information by incorporating the main privacy factors that concern people as defined by previous work [15, 56, 57, 14], and by using each cube face to deliver one (or at most two) pieces of information. This simplifies learning and provides home occupants with a sufficient understanding of the implications of IoT privacy. More importantly, because the notices are on physical device and are displayed by lighting up descriptive icons and text, PrivacyCube enables home occupants to interact with their privacy notices and gain awareness without needing a privacy background.

According to our quantitative focus group findings, PrivacyCube consistently outperforms text privacy policies in terms of comprehensiveness and usability. On the other hand, both PrivacyCube and Privacy Label were rated as comprehensive and usable. We found that the focus group participants spent more time extracting information from PrivacyCube and Privacy Label. Possible explanations include that Privacy labels were easy to navigate, PrivacyCube was new and enjoyable to explore, and participants ignored text privacy policies. Our qualitative focus group analysis revealed that participants preferred extracting information from PrivacyCube over labels and text privacy policies. Many participants found the PrivacyCube representation easier to use and comprehensive, more interactive, and more enjoyable.

Our quantitative field study indicated that PrivacyCube has an excellent usability score, and home occupants demonstrated a higher level of privacy awareness after using it. Our qualitative field study analysis also revealed that participants valued the presence of PrivacyCube in their homes as it was both decorative and functional, allowing them to discuss privacy with their families. The results also show that after a few days, participants tend to ignore PrivacyCube, but still use it if needed. This may be explained by the fact that the excitement associated with using new things decreases over time. Another important finding was that PrivacyCube encouraged more thinking about privacy and made learning privacy easier and more enjoyable for adults and children.

Overall, extracting information from PrivacyCube and Privacy Labels was perceived as a simple task and required minimal effort compared to text representations. Privacy Labels were recommended for individuals with privacy interests, and PrivacyCube was recommended for general use. PrivacyCube created an attachment and excitement in the home environment among families with children. This confirms the importance of data physicalization in creating a learning environment and suggests that PrivacyCube can improve learning about complex topics like IoT and data privacy.

Our design considerations could increase PrivacyCube's likelihood of adoption, based on our experience learning why multiple privacy interfaces have not been successful [68, 55]. According to our evaluation, PrivacyCube received high usability, increased users' privacy awareness, and improved learning and interaction with IoT privacy policies. The long field study also confirms its value in smart homes. Rather than solving the generic Web and mobile privacy interfaces issue, we present privacy notices in an appealing design and try to minimize the implications of traditional privacy policies. Considering four of the guidelines in [68, 55], we believe PrivacyCube provides a better privacy choice interface and reduces the burden of privacy-decision making for users.

### 5.2 The Need to Increase Consumers' Incentives to Consider Privacy

It is essential to consider that privacy interfaces' design can significantly influence users' decisions [55]. Despite the availability of numerous privacy interfaces, they are rarely used, and users are not motivated to consider them. Our focus group findings indicate that although participants found Privacy Labels comprehensive and usable, they were not particularly interested in using them and did not appreciate the label's medium for communicating privacy information. We argue that using data physicalization for privacy notices will nudge users into thinking more about their privacy, fostering a greater sense of awareness. By applying several aspects of the evaluation framework [55], our results show that PrivacyCube is effective in enabling users to manage their IoT privacy.

As suggested by previous research, data physicalization encourages exploration and can be an effective way to introduce individuals to computer science concepts [87]. Lederer et al. also state that better privacy designs can help individuals "understand the privacy implications of their use and engage in meaningful social interaction" [11]. Thus, physical notifications can educate users about how data is collected and used and provide an opportunity to talk about data processing with others. Data physicalization has, therefore, the potential to help individuals become more aware of privacy and spread privacy knowledge without the need to read long policies. Our findings support previous research and suggest that PrivacyCube enables individuals to craft their privacy attitude, communicate with others about privacy, and develop an engagement with IoT data privacy.

Overall, our research shows that employing data physicalization provides a scope for exploring IoT data privacy than using digital notifications. These results are consistent with other studies that indicate the cube attracts attention [88, 13]. It can be placed in the home or shared space for a group to see rather than being pushed to an individual's smartphone (like most default privacy notifications). PrivacyCube can be placed and integrated into different rooms throughout the house due to its portability, design, and color. However, its current non-waterproof material may not be appropriate for use in the bathroom. The results also show that PrivacyCube can blend into a room and appear to be another household object, such as a lampshade. These echos that PrivacyCube provides users with a neutral ground to interact with IoT privacy.

### 5.3 Limitation and Future Work

PrivacyCube has shed light on data physicalization's potential to deliver users with usable privacy notifications. Despite this, we identified some limitations in PrivacyCube.

First, PrivacyCube visualizes five privacy policy factors, displaying eight data types, access, and usage categories. Although we attempted to include more categories on each cube face, we discovered that doing so would result in a larger cube with more information, making it uninteresting to users. In addition, while we incorporated text with icons, PrivacyCube lacks explanations of what each icon and text means. Future research could tackle this issue by displaying more guidance on the screen or by including a button for more data types, usages, and access.

Second, whilst PrivacyCube monitors actual devices' status in our university's smart lab, the devices' status in our participants' homes were simulated. We opted to use simulated activities in the field study so that participants do not feel their data is invaded. This approach also allowed us to test users' privacy awareness against various IoT devices that are difficult to supply for participants, such as smart fridges and smart baths. As PrivacyCube is primarily intended to encourage learning, we believe that deploying PrivcyCube with simulated privacy notifications in people's homes for a long time provided a realistic setting. Future research could look into the PrivacyCube user experience by, for example, providing users with IoT devices to use in their homes. The programming of the cube could also be further enhanced to include a library of privacy policies, allowing for easy updates as policies change. Because such studies involve collecting personal information, they may necessitate additional considerations.

Third, participant understanding of PrivacyCube's reporting capabilities posed some limitations for the research. Although participants were informed explicitly in the information sheet and the initial meeting that PrivacyCube visualizes the behavior of specific manufacturers' privacy policies, some expressed confusion regarding PrivacyCube's data reporting. This limitation is especially pertinent given that PrivacyCube may not report the behavior of all devices in a participant's home. In addition, participants may not have all IoT devices included in PrivacyCube, or they may have different manufacturers or models of the devices. Future research could address this limitation by providing participants with the same IoT device from the same manufacturer or communicating PrivacyCube's reporting capabilities on multiple occasions to ensure comprehension.

Fourth, PrivacyCube extends our knowledge of how data physicalization privacy can help foster shared household moments. The physical aspect of PrivacyCube enabled it to seed interactions, leveraging its central position within the household and capturing attention to encourage unobtrusive engagement among the home occupants. Although we designed PrivacyCube to be a product-like design, it could incorporate interesting future enhancements. Future research could integrate audio notifications with light notifications for more modalities. On top of sending privacy notifications, PrivacyCube could also serve, for example, as a speaker to play music or as an energy efficiency monitor to display IoT energy consumption.

Our study involves a field study of 8 households with multiple members and a two-stage focus group with 6 participants. While we understand the importance of larger sample sizes, given the nature of the study, with daily observations and interactions, it was challenging to include a larger number of participants. However, this number is consistent with similar studies [83, 89, 13, 90, 91] that have employed field studies and qualitative methods. Despite our small sample, we were able to observe a variety of family interactions and arrangements. The inclusion of families with multiple members enriches our dataset, allowing us to delve deeper into the dynamics and variations within households, where each served as a microcosm, providing valuable insights.

Furthermore, the PrivacyCube design can be customized in a variety of ways. Because currently, the screen is on the cube's top face, users must exert extra effort to determine which device is active. A simple solution to this problem is to make the screen pop out so users can see it from a distance. The current rotation feature in PrivacyCube is manual, requiring users to exert additional effort to turn the cube. Future cube modifications could include an automatic rotation. Although we used high-quality wood in PrivacyCube, it is not waterproof, making it unsuitable for use in wet areas. Future research could take this into account when designing data physicalization.

Lastly, our study is primarily concerned with privacy awareness and does not explore privacy control. Providing control is challenging and not within scope, as we wanted our design to be simple, easily perceived, and relatively low-effort for users. This work suggests that data physicalization privacy will encourage people to understand the significance of the data IoT devices collect, allowing them to make informed decisions about sharing and using their data. Upon awareness, future research can explore how to provide users with privacy control interfaces.

## 6 Conclusion

We describe the design, implementation, and evaluation of PrivacyCube, a novel data physicalization to increase privacy awareness in smart home environments. PrivacyCube can be used to visualize IoT devices' data consumption and display associated privacy-related notices. We conducted a two-phase focus group study and a 14-day field study with home occupants to evaluate PrivacyCube throughout the different stages of design. Our results show that participants had a better comprehension of IoT privacy and showed an increased level of privacy awareness. Our data physicalization was also found to score high on usability and promoted non-biased privacy conversations between home occupants. Our findings indicate two potential directions for further research. The first is the use of data physicalization to promote privacy control. For instance, users can have the opportunity to configure their privacy preferences via physical means rather than using an app or web-based portal, which we believe will enable more people from diverse backgrounds to interact more confidently with IoT devices. The second is to include other uses in the privacy interface to encourage users to adopt better privacy behavior. Data physicalization can act as a more intuitive mechanism for users to improve their privacy awareness. We speculate that they may be useful for other purposes around the house or in shared spaces, such as a shared music player, a phone charging pad, and more. This study demonstrated that PrivacyCube could serve as an initial step towards a more comprehensive implementation of data physicalization privacy that enhances privacy notification visualizations within specific contexts.

## A PrivacyCube's Implementation

### A.1 Outer Casing

The final PrivacyCube contains two instances, an inner case containing the components and an outer case containing the engraved icons and texts. Both cases were laser cut from plywood (inside) and Alder wood (outside). We chose Alder wood for the outer case because it is a high-quality wood that can be used as a decorative element in people's homes. To achieve a product-like design, we covered the component holes in the outer case with laser-cut acrylic. We also painted the engraved icons and texts with flat matt black paint to make them more visible and appealing. The top and bottom faces have airflow holes to prevent the components from overheating. The cube can be easily assembled by slotting together, with no glue or tools required. The cube's components are easily accessible through the bottom face, allowing for any necessary modifications. We attached a *lazy Susan* metal to the bottom face so the cube could rotate, allowing for easy interaction.

### A.2 PrivacyCube Local Network Monitor Service

Implementation of our PrivacyCube is based on a small home network shown in Figure 12. The communication is initiated either by an IoT device in a local network or a cloud service communicating back to the IoT device. We developed a Python-based service, PrivacyCube Local Network Monitor (PCLNM), which runs on the RaspberryPi module and also acts as an internet gateway for all internal devices. The system forwards Domain Name System (DNS) queries to an external DNS server and runs a Network Address Translation (NAT) service to allow devices to communicate with the Internet. A logical architecture which demonstrates the process flow of the implementation is presented in Figure 13.

The script sniffs, using pyshark, for network packets to determine the source or destination of the local device. Once the source and destination of the packet are determined the service looks into the external IP to Country (IP2C) dataset, maintained by (ip2countrydataset). The dataset updates daily to keep the system updated, the PCLNM downloads the latest release every 24 hours in the background. On top of the IP2C database, we created a static mapping of countries with continents e.g. United States (US) and Canada (CA) falls into North America (NA). For the internal IP in the packet, the system will look into the device's details i.e. device type, device name, placement area, data types, data usage, data access and retention time. These details were gathered from the privacy policies of each device's manufacturer's website. The list of these privacy policies is listed in Table 7. As a result, the service creates a JSON object and sends it to PrivacyCube over MQTT. The JSON object format is presented in Figure 14, all data are generated from Amazon Alexa privacy policy except the *DataStorage-> Location* which is derived from network packets. The source code of our PrivacyCube's implementation is available at (GitLab).

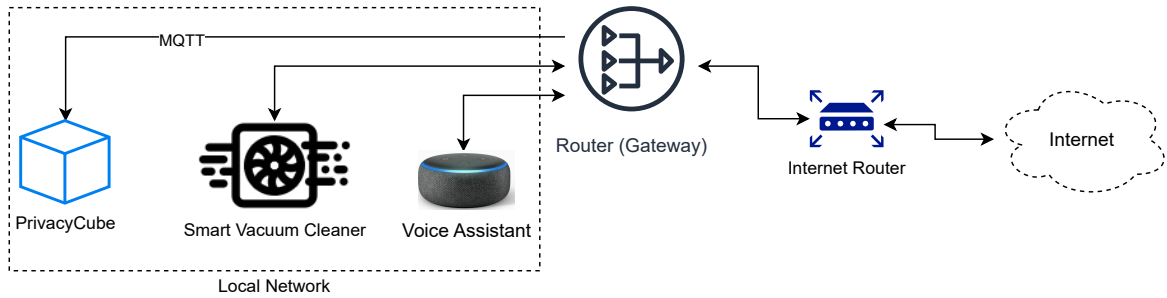

Figure 12: Physical Architecture of Implementation Setup

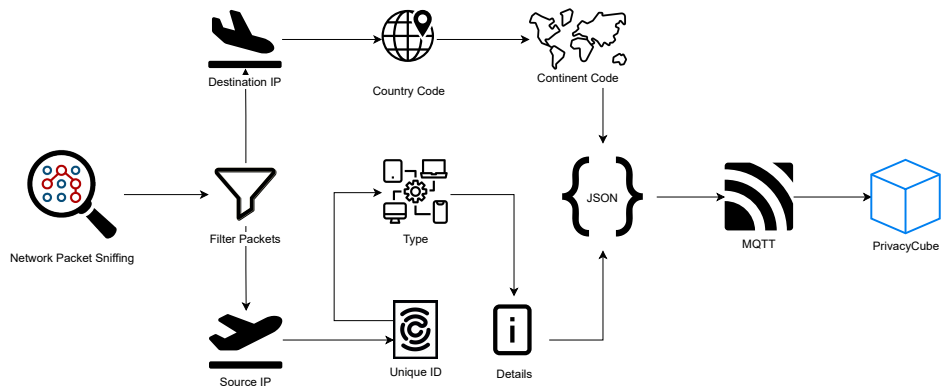

Figure 13: Process Flow of PrivacyCube System

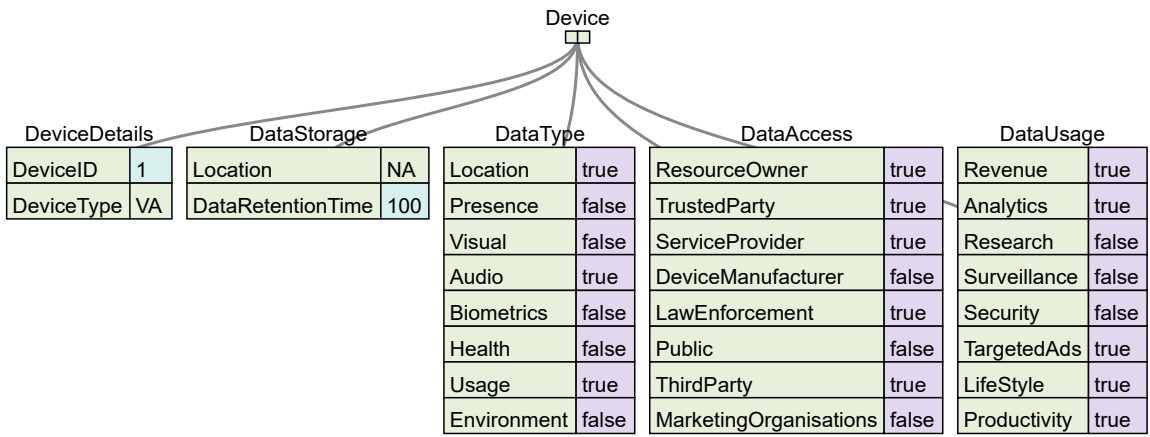

Figure 14: Data Structure of JSON File (example data for Amazon Alexa)

Table 4: Questions participants answered during the two extraction tasks.

| Extraction Tasks Questions |
| --- |
| Does the smart lock collect information about your location? |
| Does the smart lock use your location information to conduct research? |
| Does the smart lock share your information on public databases? |
| By default, does the smart lock collect your video data and use it for improving services (Reselling their services)? |
| By default, does the smart lock retain your data forever? |
| Does the sleep monitor use your health information (such as your heart rate) to conduct research? |
| Does the sleep monitor use your health information (such as your heart rate) to infer additional information about you (such as time slept, sexual intimacy)? |
| Does the sleep monitor share your contact information with third parties? |
| Does the sleep monitor share all data it collects about you with third-party service providers? |
| Does the sleep monitor sell aggregated information about you? |

### A.3 PrivacyCube User Interface

We implemented a user interface for PrivacyCube using an Arduino Leonardo microcontroller attached to a Nextion Human Machine Interface (HMI) Touch Display to handle user's commands and a Grove Shield to handle LEDs. Figure 15 illustrates the display of PrivacyCube's user interface designed by the display. The center part of the figure shows the home screen with links to rooms on the sides of the home, each room can manage up to eight devices in the current design. Each device is represented by an icon (device type) as shown in Figure 15. The device icon goes green on either tapped by user (selected by user) or being the device generating network activity.

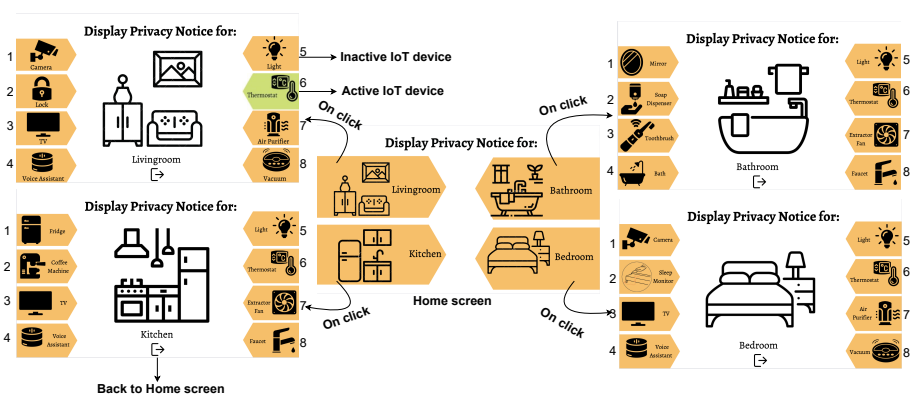

Figure 15: IoT devices are presented in PrivacyCube per home location. Each home location includes 8 IoT devices. Some devices are shared among more than one home location.

## B PrivacyCube's evaluation

Figure 16 depicts the findings of participants' responses to both the System Usability Scale (SUS) questionnaire and the adjusted user experience (UE) questionnaire. The SUS is a ten-item questionnaire with Likert-type responses on a scale of 1 to 5 (lower is better for even questions and higher is better for odd questions, 1=strongly disagree, 5=strongly agree). The UE is a four-item questionnaire on a scale of 1 to 5 (higher is better, 1=strongly disagree, 5=strongly agree).

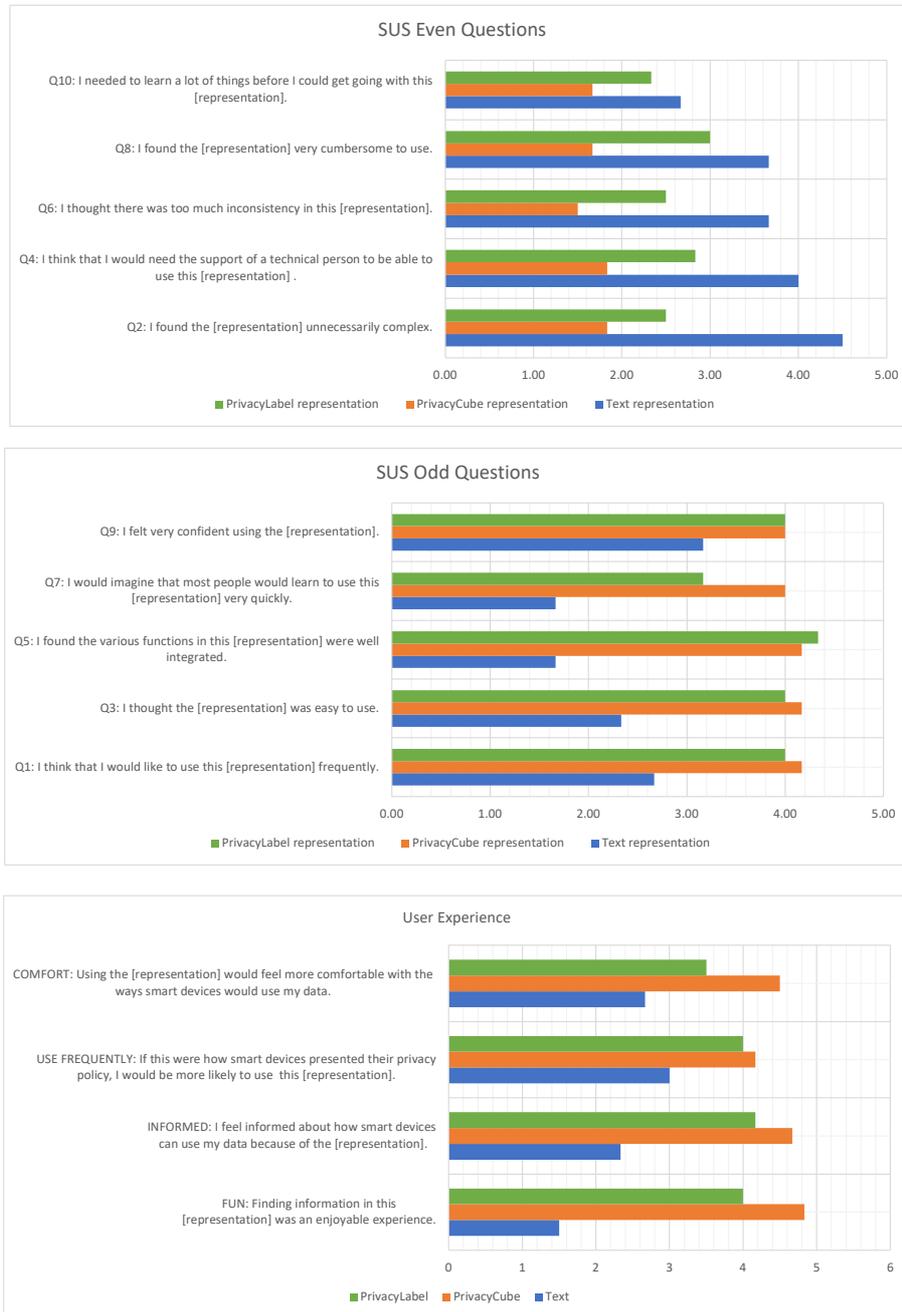

Figure 16: Participants perceived usability and user experience ratings of text, PrivacyCube, and PrivacyLabel representations. The top and middle dimensions are for SUS (out of 5, lower is better for even questions and higher is better for odd questions); the bottom dimensions are for user experience (out of 5, higher is better).

Table 5: PrivacyCube diary questions. Participants filled out a separate sheet of the same diary for each home location, i.e., living room, bedroom, kitchen, and bathroom.

| PrivacyCube Diary Questions |
| --- |
| Observing: Did you notice any IoT device sensing your data today? |
| Remembering: Did you develop any strategies to remember the IoT capability? |
| Did you know what device is on? What was it? Was it right? |
| Sharing and seeking: Did you discuss about PrivacyCube or about any IoT device with anyone today? |
| Where? Who? Did you seek out any more information? What and from what source? |
| Interaction: Did you interact with PrivacyCube? |
| Did you tap any button on PrivacyCube screen to check the IoT devices? |
| if yes, what device(s) did you check? And why? |
| Reflections: Any design suggestions/thoughts on how PrivacyCube might be different? |
| Recording: Was there any footage you would like to share? |
| Additional comments: |

Table 6: Questions we asked the participant during the second focus group study. These are guide questions, and other questions were asked based on participants' responses (probing and prompting).

| Discussion Guide Questions |
| --- |
| Did either interface leave out important information related to privacy policies? Or do you think that they were sufficient? |
| Was it easier to do the extraction task with PrivacyCube or the label? Why was it hard to use the (chosen interface)? Did you encounter any problems or obstacles while using either of the interfaces? |
| Did you have a similar level of understanding of both interfaces after using them for the same amount of time? |
| Did one interface have better or more intuitive explanations or instructions than the other? Did either interface's design or layout make it easier or harder to understand the privacy policy information? |
| Was the language used in either interface (like the icons and text) easy to understand or too technical? |
| Did either interface use animation, graphics, images, colour, or other visual or interactive elements in a way that helped explain the privacy policy information? |
| Which interface did you find more engaging or interesting to use? |
| Were there any features or functions in either interface you felt were particularly innovative or exciting? |
| Were there any specific interactions, features or functions in either product that you found confusing or difficult to use or that you still don't fully understand? |
| Which interface would you prefer to use in the future and why? |
| If you had to explain the interfaces to someone else, would you feel more comfortable doing so with one over the other? Why? |

Table 7: The IoT devices we used in PrivacyCube and their privacy policies.

| IoT device | Privacy Policy |
| --- | --- |
| Smart Light | https://www.philips-hue.com/en-us/support/legal/privacy-policy |
| Smart Thermometer | https://safety.google/nest/ |
| Smart Lock | https://en-uk.ring.com/pages/privacy-notice |
| Smart Speaker | https://aws.amazon.com/about-aws |
| Smart TV | https://www.samsung.com/uk/info/privacy/ |
| Smart Camera | https://policies.google.com/privacy |
| Smart Air-purifier | https://levoit.com/privacy-policy-for-eu-users-levoit |
| Smart Vacuum | https://global.roborock.com/pages/privacy-policy |
| Smart Fridge | https://www.samsung.com/uk/info/privacy/ |
| Smart Coffee machine | https://www.delonghi.com/en-gb/privacy-policy |
| Smart Extractor fan | https://www.gdhv.com/privacy-policy |
| Smart Faucet | https://www.moen.com/privacy-policy |
| Smart Sleep monitor | https://gdpr.jablotron.cz/documents |
| Smart Mirror | https://gesiporbath.com/policies/privacy-policy |
| Smart Bath | https://www.grohe.com/en/corporate/privacy.html |
| Smart Soap-dispenser | https://www.amazon.co.uk/gp/help/customer/display |
| Smart Toothbrush | https://shop.oralb.co.uk/information/privacy-policy |

Table 8: Questions we asked the participant after the 14-day field study. These are guide questions, and other questions were asked based on participants' responses (probing and prompting).

| Interview Guide Questions |
| --- |
| How was your overall experience? |
| Did you get annoyed by the existence of PrivacyCube in any room around your house? why? |
| Did you try to move or hide PrivacyCube if you anticipate a visitor? If yes, why? |
| Did seeing PrivacyCube let you know that there is an IoT device near you? |
| Do you get attracted by PrivacyCube? |
| Because of PrivacyCube, do you notice the functionality of the IoT devices in the vicinity? |
| After some time, do you remember the existence of an IoT device(s) around you? |
| Can you recommend any changes or adjustments to the design and functionality of Privacycube? |
| Did you change your attitude because you saw something in the device? |
| Did you stop a conversation/move from a space because the device remind you that there is a mic/camera near you? |
| Did you change your behavior after noticing the process done by the IoT devices? |
| What precautions do you take/develop while surrounded by an IoT device? |
| Do you consider yourself more confident in your privacy decision because you know the devices around you, and you know what are they collecting? |
| Does seeing what is happening to your data through the device let you change the permissions in the IoT device? |
| Did you try to obtain more information about the data collection? Did you understand what the cube is trying to tell you? |

Table 9: Participants demographics and IoT usage

| Question Type | Question |
| --- | --- |
| Demographics (asked for each household member) | Age, Gender<br>Number of Household, Data Privacy background |
| IoT usage (asked separately if more than one IoT device is used) | How many IoT devices are employed in your home?<br>Who has ownership of these devices?<br>How often do you use the devices on a daily basis?<br>For what purpose do you use the devices? |

Table 10: Awareness survey questions. Participants answered the same questions for four IoT devices, i.e., Smart Lock, Smart thermostat, Smart speakers, and Smart sleep monitor.

| I expect the (device name) to collect this data (please choose all that apply) |
|---|
| • Location |
| • Presence |
| • Visual |
| • Audio |
| • Biometrics |
| • Health |
| • Usage |
| • Environment |
| • Other (If you selected Other, please specify:) |
| I expect the(device name) to share my data with these parties (please choose all that apply) |
| • Resource Owner |
| • Trusted Party |
| • Service Provider |
| • Device Manufacturer |
| • Law Enforcement |
| • Public |
| • Third-Party |
| • Marketing Organisation |
| • Other (If you selected Other, please specify:) |
| I expect the (device name) to use my data for these purposes (please choose all that apply) |
| • Revenue |
| • Analytics |
| • Research |
| • Surveillance |
| • Security |
| • Targeted Ads |
| • Lifestyle |
| • Productivity |
| • Other (If you selected Other, please specify:) |
| I expect my (device name) data to be stored in these locations (please choose all that apply) |
| • Within device |
| • Within multiple devices |
| • Local server |
| • Remote Server |
| • Third-party server |
| • Public server |
| • Other (If you selected Other, please specify:) |
| I expect my (device name) data to be retained\collected (please choose all that apply) |
| • Indefinitely |
| • 1 Month |
| • 3 Months |
| • 1 Year |
| • Event-based |
| • Every Second |
| • Every Hour |
| • Every Day |
| • Other (If you selected Other, please specify:) |